# Electric field-induced interfacial instability in a ferroelectric nematic liquid crystal

Marcell Tibor Máthé[1,2], Bendegúz Farkas[1,2], László Péter[1], Ágnes Buka[1], Antal Jákli[1,3,4,*],
Péter Salamon[1,*]

[1]Institute for Solid State Physics and Optics, Wigner Research Centre for Physics, P.O. Box
49, Budapest H-1525, Hungary

[2]Eötvös Loránd University, P.O. Box 32, H-1518 Budapest, Hungary

[3]Materials Sciences Graduate Program and Advanced Materials and Liquid Crystal Institute,
Kent State University, Kent, Ohio 44242, USA

[4]Department of Physics, Kent State University, Kent, Ohio 44242, USA

*: Author for correspondence: salamon.peter@wigner.hu

## Abstract

Studies of sessile droplets and fluid bridges of a ferroelectric nematic liquid crystal in externally applied electric fields are presented. It is found that above a threshold, the interface of the fluid with air undergoes a fingering instability or ramification, resembling to Rayleigh-type instability observed in charged droplets in electric fields or circular drop-type instabilities observed in ferromagnetic liquids in magnetic field. The frequency dependence of the threshold voltage was determined in various geometries. The nematic director and ferroelectric polarization direction was found to point along the tip of the fingers that appear to repel each other, indicating that the ferroelectric polarization is essentially parallel to the director. The results are interpreted in connection to the Rayleigh and circular drop-type instabilities.

## I.   Introduction

Conventional nematic liquid crystals are usually formed by elongated organic molecules. In a continuum description, the local symmetry axis of nematics is defined by the average molecular orientation, which is called the director $(\hat{n})$[1]. Nematic liquid crystals have already achieved significant impact in our modern life as being the bases of the currently dominant display technology. In flat display technology, where dielectric nematic liquid crystals are utilized, the high frequency (AC) electric field-induced director reorientation results in the electro-optical effect allowing to control the brightness of each pixel in a display[2]. An important property of the nematic phase is the "head–tail" symmetry, i.e., $\hat{n} = -\hat{n}$. This head–tail symmetry is broken in the ferroelectric nematic ($N_F$) phase due to the spontaneous polarization $\vec{P_s}$, where the director becomes a vector $\vec{n}$ that is assumed to be parallel to $\vec{P_s}$ [3,4]. The existence of a ferroelectric liquid nematic phase was proposed already in 1916 [5,6] by Max Born to explain the isotropic-nematic phase transition. Although it turned out that ferroelectricity is not needed for the existence of a nematic phase, scientists were looking for the $N_F$ phase over a century, but there were no unambiguous experimental indications of it until the syntheses of the highly polar rod-shaped compounds referred to as DIO and RM734 by Nishikawa et al.[7] and Mandle et. al.[8,9], respectively in 2017. The $N_F$ phase of RM734 was first suggested to have splayed polar order[10–12], but more recently, it was shown that it has a uniform ferroelectric nematic phase[3]. RM734 and DIO have large molecular dipole moments of about 10 D, and a spontaneous polarization up to $|\vec{P_s}|{\sim}5$ μC/cm². The apparent dielectric permittivity and its anisotropy were reported to be orders of magnitude higher than those of classical nematics, reaching $\varepsilon \sim 10^4$ or higher[7,11,13–18] implying an extremely large sensitivity of the $N_F$ materials to electric fields. Such a large dielectric constant suggests a giant dielectrowetting at unprecedentedly low voltages, since its threshold scales with the square root of the fluid's permittivity[19–21]. Furthermore, in analogy to the spectacular Rosensweig instability[22] of ferromagnetic fluids in magnetic field, one can expect to see electric field induced spike patterns in ferroelectric fluids with free surface. A hint of such instability was demonstrated by Barboza et al.[23] who showed that sessile droplets of RM734 become unstable and disintegrate through the emission of fluid jets when they are deposited on a lithium niobate (LN) ferroelectric crystal substrate. This phenomenon was explained in analogy to the Rayleigh instability[24] of charged fluid droplets. Recent studies presented the behavior of sessile ferroelectric nematic droplets on LN surfaces exposed to light[25–28]. We note that ferroelectric nematic droplets in various environments[29–32] exhibit an extraordinary tangential arrangement of the spontaneous polarization.

In this paper, we reveal the nature of an AC electric field induced interfacial instability in ferroelectric nematic sessile droplets and liquid bridges that resembles Rosensweig-type instabilities observed in ferromagnetic liquids and Rayleigh-type instability observed in charged droplets. In addition to quantitative measurements in various geometries, we will also theoretically analyze the results.

## II. Experimental results

We have studied the effect of AC electric fields in three different geometries: electric voltage applied normal to the base plane of a sessile droplet (G1), and of a fluid bridge (G2), and electric field applied along the base plane of a sessile droplet (G3). Unless it is indicated differently, all measurements were carried out about 5 K below the $N - N_F$ phase transition. The phase transition temperature between the nematic and ferroelectric nematic phase was $T_{NN_F} = 408.15$ K (132 °C).

The experimental geometry of G1 is shown in Figure 1a. Micrographs of a droplet exposed to rms voltages $U = 210$ V, 220 V, and 300 V are seen in Figure 1b, 1c, and 1d, respectively. The AC voltage at $f = 10$ kHz frequency was applied between the base plate and another ITO coated plate placed at $L = 150$ μm distance. At and below 210 V the droplet is stable, while above it, the ejection of several thin jets can be observed. Some of them are decorated with additional secondary jets like that observed by Barboza et al[23]. The structure is stationary at a given voltage (see Supplementary Video 1).

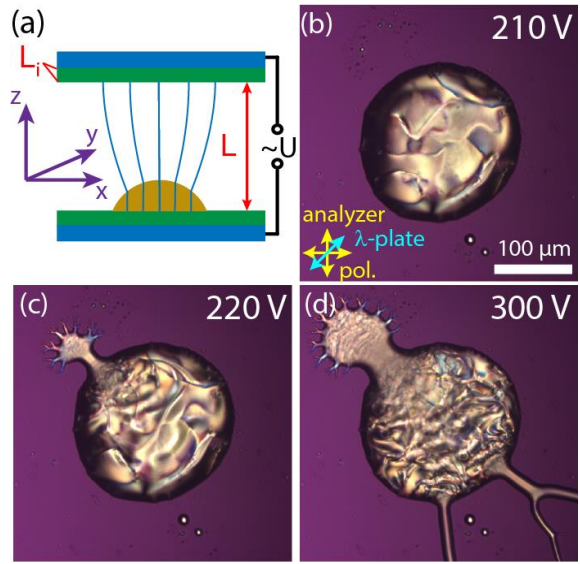

*Figure 1: A ferroelectric nematic sessile droplet in electric fields along the normal direction of the base plate. (a) Illustration of the G1 geometry with approximate electric field (green lines) and coordinate system. (b-d) Polarizing optical microscopy images of a ferroelectric nematic sessile droplet between crossed polarizers (yellow arrows) and a full-wave plate (λ=546 nm - blue arrow) at 210 V, 220 V and 300 V, respectively. The cell gap was L=150 μm.*

The G2 geometry and a typical response to electric fields of an $N_F$ bridge with $L = 74$ μm thickness is shown in Figure 2. At high temperatures, in the isotropic and the nematic phase, the circumferences of the contact lines on the bottom and top bounding plates start to grow above a certain threshold voltage, while keeping their original round shape (Figure 2b). In contrast to this behavior, in the ferroelectric nematic phase (Figure 2c) the contact line becomes unstable and a fractal-like spreading of the fluid is observed (see Supplementary Video 2 for another droplet). Inside the droplet at high voltage, electro-hydrodynamic convection takes place that is stronger near the perimeter of the droplets. To precisely analyze the voltage dependence of the electric field induced interfacial instability, we applied a geometrical transformation, mapping a band around the contact line to a rectangle (Figure 2d), which we will further discuss below. In Figure 2e, we show the side view of a liquid bridge exposed to high voltage, clearly demonstrating that the instability forms in the vicinity of the glass plates. We note that the visibility of the upper surface in the image in Figure 2e is blocked by the edge of the upper substrate due to the inclined axis of observation.

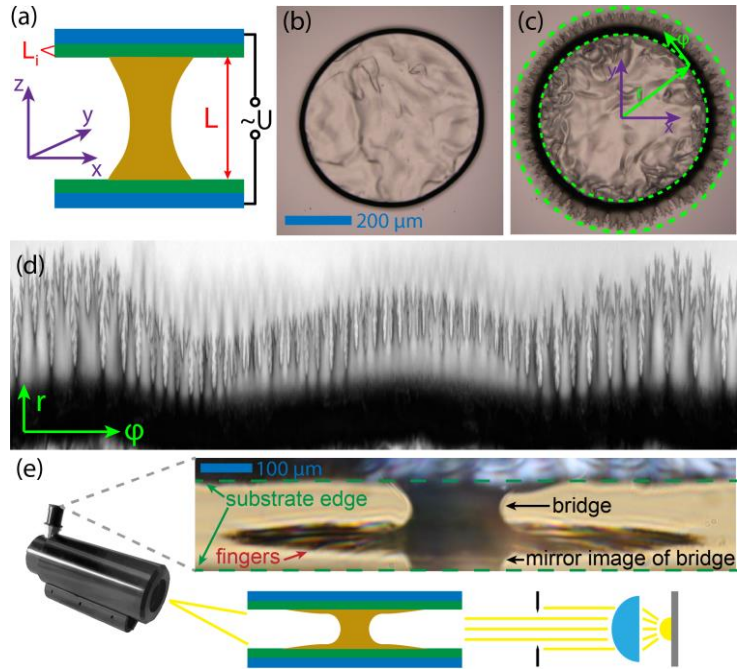

*Figure 2: A ferroelectric nematic fluid bridge in electric field along the normal direction of the base plate. (a) Illustration of the G2 geometry. (b) Top view of a fluid bridge at zero voltage (558 µm diameter, 74 µm thickness) and (c) at f =800 Hz, 52 V AC voltage. (d) Transformed view of the perimeter of the bridge exhibiting the instability. Darker (lighter) pattern shows the instability at the top (bottom) glass. (e) Side view of the bridge clearly indicates that the instability forms in the vicinity of the electrodes. The schematic illustration describes the experiment for side-view imaging using a long-range microscope.*

Using the transformed image of the perimeter of the droplet, we calculated a specific type of roughness of the contact line denoted by $Ro_{cl}$ and defined in detail in the Methods section. $Ro_{cl}$ gives a quantitative indicator for the emergence of the interfacial instability, while being insensitive to the gradual growth of the circumference due to electrowetting.

Figure 3 shows the schematics of the in-plane electrode geometries (G3) together with microphotographs of ferroelectric nematic droplets. Figure 3a illustrates the interdigitated electrodes with the coordinate system, while Figure 3b and c show the side view ($x − z$ plane) of the electrodes with the fringing field, and the sessile droplet, respectively. The top-view ($x − y$ plane) of the reflection microscopy images of a droplet at zero voltage and with $f = 10$ kHz, $U = 75$ V applied voltage are shown in Figure 3d and e. The latter case is just above the threshold of the interfacial instability, and the emergence of the branched fingers is observable. Supplementary Video 3 also illustrates the effect. The protruded parts of the ferroelectric liquid are located on the brighter stripes corresponding to electrode areas as shown in comparison with Figure 3c. Note, that here both the electrode width and gap distances are 10 μm. Figure 3f illustrates the side view of the electrodes where the gap distance (180 μm) is much larger than the electrode width (10 μm). Figure 3g shows the corresponding polarizing optical microscopy (POM) image of a sessile droplet above the threshold of the instability with a full-wave plate inserted at 45° with respect to the crossed polarizers. The image clearly shows that the fingers are located on top of the electrodes, where the fringing field is mostly perpendicular to the substrate (see also Supplementary Video 4).

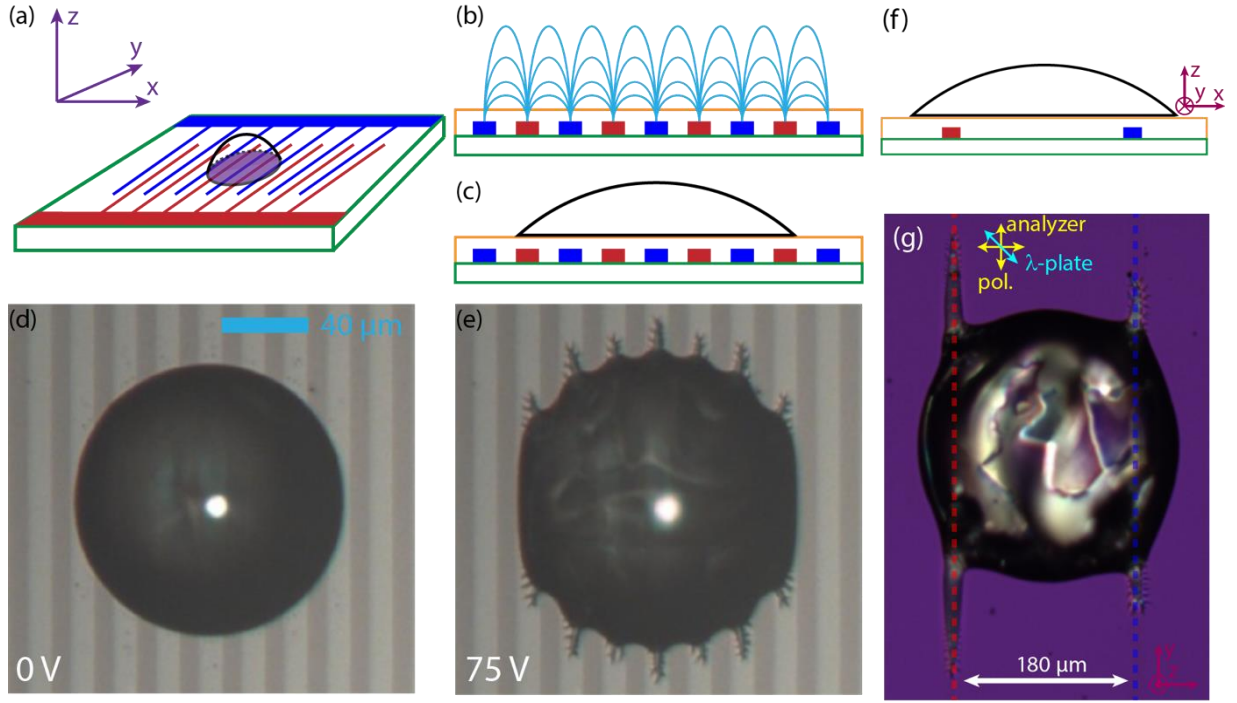

*Figure 3: Schematics of the in-plane (G3) electrode geometry and micrographs of the ferroelectric nematic droplet on interdigitated electrodes. (a): Illustration of the interdigitated electrodes with the coordinate system. Side view (x − z plane) of the electrodes with fringing field (b) and the sessile droplet (c). Reflection microscopy image (x − y plane) of the droplet at 0 V (d) and 75 V (e) applied voltage at f = 10 kHz. Brighter stripes correspond to the electrode areas; the electrode distance was 10 μm. (f) Side view of the geometry with a single thin pair of electrodes with a larger gap (180 μm) between them. (g) The corresponding polarizing microscope image, showing that the fluid fingers spread above the electrodes, where the fringing field is perpendicular to the substrate.*

Figure 4 shows the temperature and frequency dependences of the threshold voltages for geometries G2 and G3. Figure 4a shows the temperature dependence of the threshold voltage $U_{th}$ at $f = 1$ kHz, revealing that the emergence of the contact line instability occurs at much lower voltages on cooling from the $N$ to the $N_F$ phase. Note, that the transition temperature $T_{NN_F}$ corresponds to the state without voltage. It is known[33] that the electric field might shift the transition by several degrees. In the ferroelectric nematic phase, the threshold voltage slightly decreases towards lower temperatures. The small, but clearly distinguishable hysteresis in the voltage dependence of the contact line instability is present in the entire temperature range. The inset of Figure 4a shows $Ro_{cl}$ as function of the applied voltage at $f = 1$ kHz and $T = 400.15$ K (127 °C) ($T - T_{NN_F} \approx -5$ K) in the ferroelectric nematic phase. Up to 42 V the roughness is constant in spite of the slight increase of the circumference of the contact line; above 42 V a sharp increase of the roughness clearly indicates the appearance of the spiky structure, and the threshold-like behavior. Decreasing the voltage, a similar behavior is observed with a threshold at 38 V. This weak hysteresis indicates a weakly first order nature of the instability. Quasi-statically increasing the voltage led to the same pattern as when the voltage was suddenly switched above the threshold from zero.

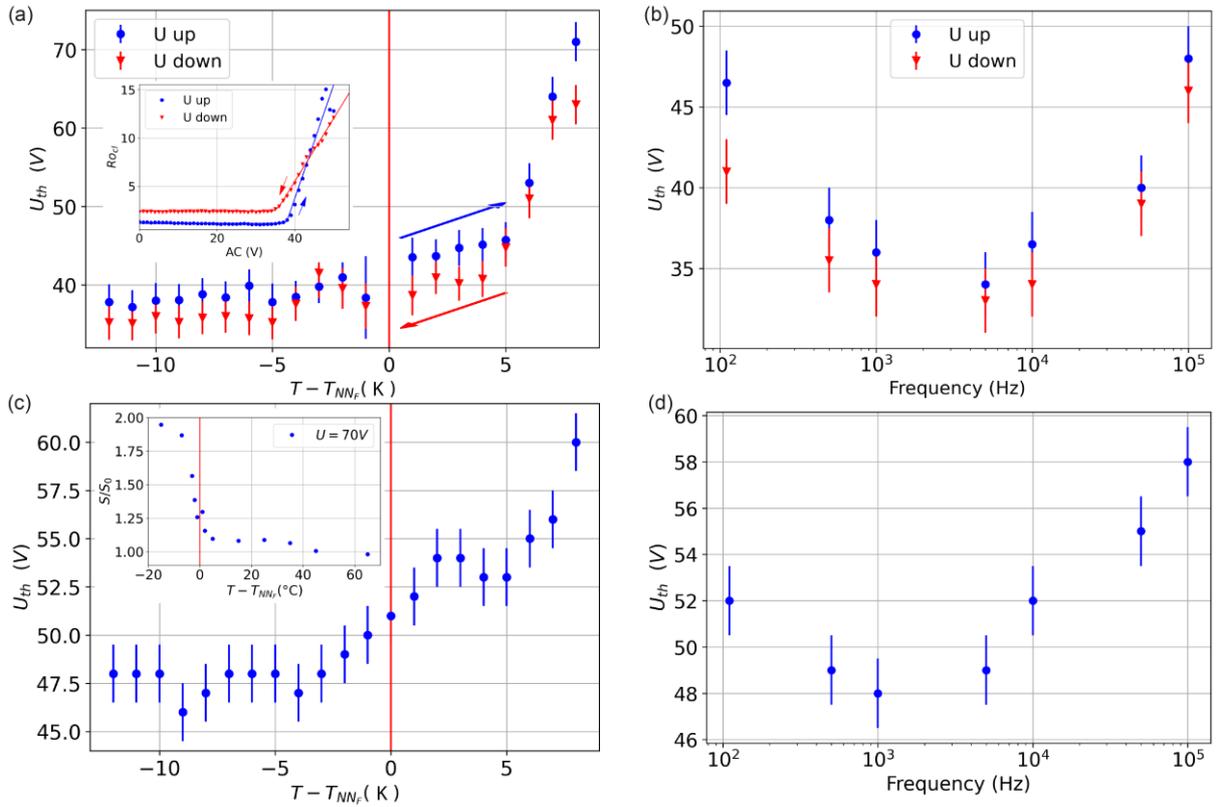

*Figure 4 (a) Temperature dependence of the threshold voltage measured at f = 1 kHz in the fluid bridge geometry (G2) with cell thickness L = 74 μm. (a-inset) The roughness of the contact line (Ro_{cl}) of the fluid bridge at one substrate as function of applied voltage at T = 400.15 K (127 °C) (T − T_{NN_F} ≈ −5 K) and f = 1 kHz. (b) Threshold voltage U_{th} of the interfacial instability as function of frequency at T = 400.15 K (127 °C). The threshold voltage of the interfacial instability measured on substrates with interdigitated surface electrodes (G3 geometry, electrode distance: 10 μm) (c) as a function of temperature at f= 1 kHz and (d) as a function of frequency at T = 400.15 K (127 °C). (c-inset) The normalized circumference S/S_0 of a 125 μm diameter droplet as a function of temperature with a constant applied voltage (U=75 V, f =10 kHz).*

In Figure 4b, we have plotted the frequency dependence of the threshold voltage in the ferroelectric nematic phase at $T = 400.15$ K (127 °C) ($T - T_{NN_F} \approx -5$ K). In the $f =$ 100 Hz – 100 kHz range the lowest thresholds were found at around $1 \text{ kHz} - 5 \text{ kHz}$ and $U_{th}$ sharply increases toward the low and high ends of the studied frequency range. The hysteresis between decreasing and increasing voltages was found to be larger at the lower and higher limits in the studied $f$-range.

The temperature dependence at $f = 1 \text{ kHz}$ and the frequency dependence at $T = 400.15 \text{ K } (127 \text{ °C})$ of the threshold voltage of the interfacial instability are shown in Figure 4c and Figure 4d, respectively. The temperature dependence at 1 kHz shows a gradual decrease of the threshold voltage while cooling from the $N$ to the $N_F$ phase. Cooling in a constant voltage, the normalized circumference remains near one in the isotropic and the upper part of the nematic range, but it increases sharply already in the $N$ phase before entering to the $N_F$ phase. The inset to Figure 4c shows the temperature dependence of the normalized circumference $S/S_0$ of a 125 μm diameter droplet for $U = 75 \text{ V}, f = 10 \text{ kHz}$, where $S_0$ is the circumference of the drop at the base substrate without applied voltage. One can see that the threshold voltage in the $N_F$ phase has a minimum at around 1 kHz. Interestingly, as seen in Figure 4, the numerical values of $U_{th}$ do not differ much in the completely different geometries of G2 and G3.

Figure 5 shows the electrically induced pattern on wide in-plane electrode pairs, where the electrode width was much higher than the initial droplet diameter and the separation of electrodes. The gap between the electrodes was 180 μm and the diameter of the drop was always larger than the gap. As we increase the applied voltage in this geometry, the drop starts to spread onto the surface of the electrodes, where the electric field has a dominant z-component. During the spreading, several fingers start to grow from the drop evolving to a fractal-like structure as seen in Figure 5c (see also Supplementary Video 5). This effect increases up to 90 V, then at higher voltages the drop starts to spread parallel with electrodes, too (Figure 5d). By applying even higher voltages, the drop continuously spreads along the electrodes and becomes increasingly elongated (see also Supplementary Video 6). It is important to emphasize that the branched structure does not change in time as long as the applied voltage is kept constant. When a series of droplets are deposited in a line between the electrodes, we find that in case of the developed instability the drops are not merged but they repel each other as seen in Figure 5e. A closer look at the POM image taken with a λ-plate sheds light on the molecular arrangement in the branches. In Figure 5f, we can see blue and orange tips meaning parallel and perpendicular orientation of the director with respect to the axis of the λ-plate, respectively. This observation indicates that during the finger growth, the director and thus the spontaneous polarization are aligned along the branches. This explains the electrostatic nature of the repulsion between neighboring fingers facing each other.

Using an image of a well-developed branched structure in the geometry of wide in-plane electrodes, we have determined the Hausdorff or fractal dimension as $d_f \approx 1.61 \pm 0.1$, which indicates a fractal-like property of the structure since $d_f < 2$ .[34]

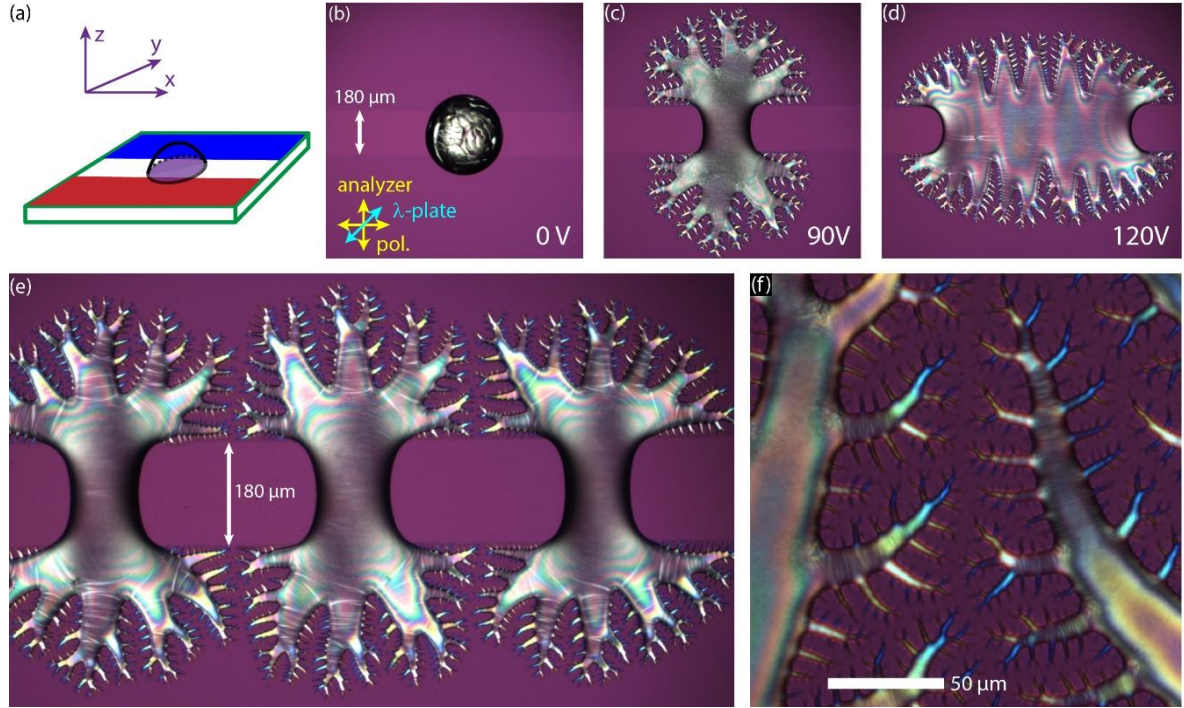

*Figure 5: Geometry of the wide in-plane electrodes (a). Polarizing optical microscopy images of ferroelectric nematic droplets on surface electrode pairs (with 180 µm distance) at 0 V (b), at 90 V (c) and at 120 V (d) with f = 500 Hz sinusoidal driving. (e) A series of neighboring droplets showing the instability and interaction. (f) A magnified area near two branches showing that their tips are arranged to avoid contact between neighbor tips.*

To analyze the director structure more quantitatively, we carried out polarimetric microscopy measurements of the tips giving pixelwise information of the average director orientation in the sample plane. In Figure 6a, a typical example is presented, where the map of green rods represents the projection of the director field on the sample plane. The map clearly shows that the director is aligned along the spreading direction of the fingers, which leads to bound charges at the surface that repel each other. We also detected the surface topography of the branched structure in crystallized samples quenched from the ferroelectric nematic phase. A typical result from a white light interferometric profilometer is shown in Figure 6b, where the color-coded height profile is measured from the flat base substrate. We selected one finger from the lowest level of the hierarchical branched structure (see gray dashed line in Figure 6b), and plotted the height profile of it in a cross-sectional view as seen in Figure 6c. We find that the width and height of the finger are about 2 µm and 0.5 µm, respectively. Such dimensions are close to the resolution of our optical microscopy techniques; therefore, we applied scanning electron microscopy (SEM) to reveal the fine structure and tip dimensions the electric field induced fingers using quenched samples. A representative SEM image is shown in Figure 6d, where the branched structure can be observed with finer details. The magnified view of a selected tip presents the submicron radius of curvature (~0.5 μm) at the end of the finger. We note that the profilometry and the SEM experiments were performed on the same area of the same sample.

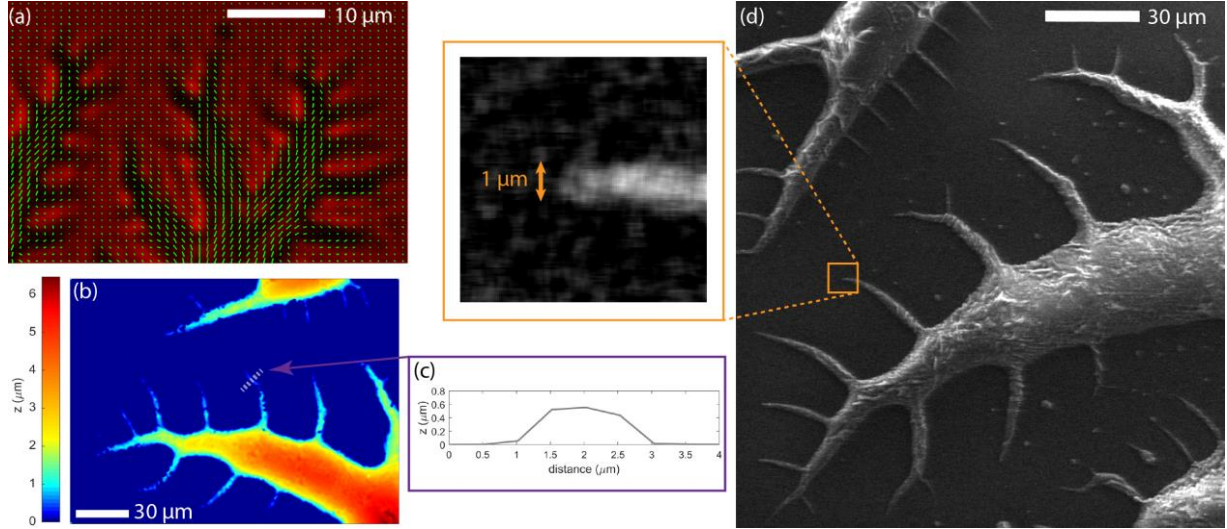

*Figure 6: Properties of the tips of the side branches of the ferroelectric nematic liquid crystal sessile droplet above the electrically induced fingering instability. (a) Polarimetric microscopy image of the branched ferroelectric fluid interface on wide surface electrode pairs. Green rods indicate the effective orientation of the director. (b) Surface topography of the branched structure measured by interferometric profilometry. (c) Cross-section of one branch indicated by a dotted gray line. (d) Scanning electron microscopy image of the quenched ferroelectric nematic fluid exhibiting the electric field-induced interfacial instability.*

## III. Discussion

Our results presented above reveal an AC electric field-induced fingering instability resembling to the Rosensweig[22] or normal field instability, which is a well-known interfacial pattern formation phenomenon observed in ferromagnetic fluids, where the horizontal ferrofluid surface becomes unstable above a critical magnitude of vertical magnetic field producing spectacular spiky extrusions[22,35–37]. Depending on the geometry, such instability can manifest itself in various morphologies, but the basic mechanism was understood as a balance between gravitational, capillary, and magnetic forces. While Rosensweig instability produces spikes normal the free surface, in our case they appear on the bounding plates normal to the electric field. We think the main reason for the difference is that for our small sample sizes the effect of gravity is negligible compared to the surface tension.

A fingering or labyrinthine instability driven by a magnetic field perpendicular to the plates of thin films or the base plates of small droplets where the effect of gravity can be neglected[38–41], is more analogous to our experiments[42]. Such labyrinthine instability was also observed in dielectric fluids exposed to several kilovolts[35,38,43]. In both types of fields, a required condition for the emergence of the labyrinthine instability is a gap between the ferromagnetic fluid and the poles of the magnet or between the dielectric fluid and the electrode surface. Due to the presence of such a gap, a field component parallel to the plane of the bounding substrates can arise. Our preliminary results in the liquid bridge geometry clearly indicate the emergence of the labyrinth instability at a much higher threshold voltage than required for the reported interfacial branching (ramification). That secondary instability of the contact line can undoubtedly be associated to the labyrinth instability because of the specific morphology identical to the previous findings in magnetic and dielectric fluids. The analysis of this secondary instability is out of the scope of this paper and will be the subject of a further work.

For the liquid crystal we studied here, the appearance of the electric-field-induced ramification phenomenon can be also seen in the paraelectric nematic phase[44,45], but at much higher threshold voltages that sharply increase toward higher temperatures, as seen in Figure 4. This indicates that the required polarization should not necessarily originate from a spontaneous ferroelectric polarization but a large induced polarization could be similarly effective. In fact, a recent study[33] showed the induction of ferroelectric nematic order even 30 K above the $N - N_F$ phase transition. Note that in the isotropic phase of RM734 and in the nematic fluid of 5CB we found no ramification.

The results in geometry G1, when external electric field was applied normal to the base plate of sessile droplets, are very similar to that observed[23,25,26] when the sessile droplets were placed on a lithium niobate ferroelectric crystal surface, indicating a Rayleigh-type instability in which droplets exhibit the sudden emission of fluid jets with electric charging beyond a threshold. In those ferroelectric sessile droplets, the bulk liquid crystal polarization $P_s$ spontaneously self-organizes in response to the external electric field provided by the fringing electric field of the ferroelectric crystal.

Our experiments differ in several respects. (i) Our droplets and bridges do not sense the DC fringing field of a ferroelectric crystal, but we apply external AC fields either normal to the base plate or in-plane. (ii) AC field induced effects can also be seen in bridges (G2) and in-plane (G3) geometries. Despite these differences, for all geometries the instability observed along the base substrate is due to a field that is perpendicular to the substrates just as in Ref. [23].

In all studied geometries, the threshold voltage of the instability sharply increases towards lower frequencies as seen in Figure 4(b) and Figure 4(d). Decreasing the frequency below 100 Hz, the threshold voltages are more and more difficult to measure due to higher error and less reproducibility. At DC, the threshold field is above the electric breakdown of the sample. Depending on the magnitude of the driving electric field, the switching time of the spontaneous polarization is between 10 µs and 1000 ms [46], which corresponds to the frequency range 100 kHz – 1 kHz. Accordingly, in the studied frequencies the polarization can follow the AC fields, thus the AC field induced effect can be modelled in analogy to the DC effect of a ferroelectric crystal. The main advantage of the AC field induced phenomenon, in addition to the more versatile geometries, is that the screening effect of mobile ionic contaminants is smaller.

In G1 geometry (Figure 1a), assuming the liquid crystal droplet is perfect insulator, the internal electric field $E_i$ is scaled down by the ratio of the dielectric constant of the droplet $\varepsilon_d$ and the surrounding medium $\varepsilon_m$ as $\vec{E}_i = \frac{\vec{E}_o}{1+N(\varepsilon_d/\varepsilon_m-1)}$, where $N$ is the depolarization factor. For spheroid with the rotational axis $c$ being along the field, $N$ depends on the ratio $c/a$, where $a$ is the radius at the base plate. Specifically, when $c << a$, then $N \approx 1$ and $\vec{E}_i \approx \frac{\vec{E}_o \cdot \varepsilon_m}{\varepsilon_b}$ (see [47]). It has been shown recently that for a hemispheroid the internal electric field is very close to that of a spheroid[48]. Approximating our spherical cap with a shallow hemispheroid, we obtain that the internal field in the liquid crystal sessile droplet is roughly inversely proportional to the effective dielectric constant, which is close to $\varepsilon_\perp$, i.e., $E_i \approx \frac{E_o}{\varepsilon_\perp} = \frac{U}{L\varepsilon_\perp}$.

Assuming that the induced polarization $\Delta P$ will compensate the internal field $E_i = -\frac{\Delta P}{\varepsilon_o \varepsilon_\perp}$, we get that $\Delta P \approx \frac{U \cdot \varepsilon_o}{L} \sim 10^{-5}$ C/m$^2$, i.e., the same order of magnitude as estimated by Barboza et al.[23] for droplets on ferroelectric crystal plates.

Following the arguments in reference [[23]], we assume that the ramification instability starts in a topological defect with surface area $S$ and charge accumulation of $q = 2P_s S$. The condition for the finger formation is that the repulsive Coulomb force $F_C = \frac{kqQ}{l^2}$ overcomes the stabilizing force arising from the surface tension $F_S \sim \gamma \pi r$ at a protruding tip with an effective radius of curvature $r$. At the onset of the instability, the distance $l$ between the effective charges equals the base radius of the droplet, i.e., $l = R$. With $k = \frac{1}{4\pi\varepsilon_0}$, $q = P_s \pi r^2/2$ and $Q = \Delta P \cdot R^2 \pi$ and $\Delta P \approx \frac{U \cdot \varepsilon_o}{L} = E_o \varepsilon_o$ we get the condition for the external electric field $E_o$ that $E_o > \frac{2\gamma l^2}{\pi k P_s r \varepsilon_o R^2} \geq \frac{2\gamma}{\pi k P_s r \varepsilon_o}$. The threshold voltage is approximated by

$$U_{th} \approx \frac{2\gamma L}{\pi k P_s r \varepsilon_o} = \frac{8\gamma L}{P_s r}. \qquad (1)$$

At 5 K below the $N - N_F$ transition, the spontaneous polarization was measured[3] as $P_s \approx 4 \cdot 10^{-2}$ C/m$^2$ and the surface tension of RM734 was estimated[23] to be $\gamma \approx 0.01$ N/m. As seen in Figure 1c and Figure 6(c-d), the curvature radii of the tips are about $r{\sim}1$ μm. Substituting the above parameters and the electrode distance $L = 150$ μm, eq.(1) results in $U_{th} \approx 300$ V, which is close to that found in the experiments despite the simplicity and the approximations of this model.

In summary, we presented pattern forming instabilities of sessile droplets and fluid bridges of a ferroelectric nematic liquid crystal in externally applied electric fields. The fingering (or branching) instability observed above is similar to the Rayleigh-type instability observed in charged droplets in electric fields. The nematic director and ferroelectric polarization direction was found to point along the tip of the fingers that appear to repel each other, indicating that the ferroelectric polarization is essentially parallel to the director.

## IV. Methods

In our studies, we used (4-[(4-nitrophenoxy)carbonyl] phenyl-2,4-dimethoxybenzoate (RM734), one of the prototype compounds for ferroelectric nematic materials [4,8,9]). The sessile liquid crystal droplets were prepared in the nematic phase (at ~ 443.15 K (170 °C)) by a custom-made setup allowing simultaneous microscopic observations in both horizontal and vertical directions by applying two Questar QM100 long-range microscopes with a Canon EOS450D and a Hayear HY-500B camera. According to our experience, the preparation of ideal, circular shaped droplets is easier at higher temperatures, however, keeping the compound RM734 at temperatures higher than 443.15 K (170 °C) for longer times may lead to the degradation of the material, which we aimed to avoid. Positioning of droplets was done in a heated environment (with temperature stability better than 0.1 K) by a micromanipulator arm. The experiments on the droplets were carried out by using a Leica DMRX polarizing microscope equipped with a Linkam LTS350 hot stage and a TMS94 controller providing a temperature stability of 0.01 K.

In all our sample geometries we have used either plain or patterned indium-tin-oxide (ITO) coated glass plates as electrodes, which were coated by a 1.4 μm thick layer of SU8-3000 (Kayaku Microchem) hard baked photoresist to prevent direct charge transfer. The unrubbed SU8 layer provided degenerate planar alignment for the liquid crystal. We note that there is a significant effect of the thickness of the SU8 layer on the electrodes due to its effect on the voltage distribution along z. In case of the liquid bridge geometry, the threshold voltage for the Freedericksz transition in the nematic phase of RM734 was 40 V with 1.4 μm thick SU8 coatings on the electrodes, while for 260 nm thick insulator layer, this threshold was found to be as low as 1 V. The contact angle of RM734 on SU8 was found to be $\theta_c \approx 41° \pm 5°$.

For scanning electron microscopy and optical profilometry measurements, the sample in the $N_F$ phase was quenched by suddenly dropping it into liquid nitrogen while the electric field was applied onto it. Rapid solidification preserved the fine branched structure of the ferroelectric nematic fluid. Scanning electron micrographs were recorded with a TESCAN MIRA3 scanning electron microscope equipped with a field-emission gun. Since the solidified liquid crystal samples are electrically insulating, the low-vacuum mode was applied with 40 Pa nitrogen pressure to suppress the charging of the sample. Under these circumstances, images with sufficient resolution could be obtained in the 5-8 kV acceleration voltage range. An LVTSD detector was used to obtain secondary electron images. Optical profilometry measurements were performed with a Zygo NewView 7100 3D white light interferometric profiler using Mirau interferometric objectives. Polarimetric microscopy measurements were performed by applying a Thorlabs Kiralux CS505MUP polarization camera in a custom-made microscope with circularly polarized monochromatic illumination (at 660 nm wavelength) in transmission mode.

To precisely analyse the voltage dependence of this instability, we applied a geometrical transformation, mapping a band around the contact line to a line (Fig2 (c)). To transform the surface of the drop to a straight line, first the undeformed drop was fitted by an ellipsoid, then the images close to the surface of the drop were plotted. We define the roughness of the contact lines for a given voltage as $Ro_{cl} = \sum_\varphi \left| \partial_\varphi^2 \sum_r I(r,\varphi) \right|$, where $I(r,\varphi)$ is the pixel intensity map of the image as a function of the radius $r$ and the azimuthal angle $\varphi$. $\partial_\varphi^2$ denotes the second derivate and the sum was calculated in the surroundings of the contact line. The voltage dependence of the roughness of the contact lines was fitted by the function[49] $f(U) = a \cdot \ln\left(1 + e^{b/e \cdot (U-U_{\text{th}})}\right) + d$, where $U_{\text{th}}$ is the threshold voltage, $a$, $b$, and $d$ are fit parameters.

The fractal dimension was determined by image analysis applying the box counting algorithm implemented in the PoreSpy[50] module of Python.

## Acknowledgement

This work was financially supported by the Hungarian National Research, Development, and Innovation Office under grant NKFIH FK142643 and the US National Science Foundation under grant DMR-2210083. This paper was supported by the János Bolyai Research

Scholarship of the Hungarian Academy of Sciences. We are thankful to Ewa Körblova and David Walba at University of Colorado at Boulder for providing RM734 for us. We thank Attila Nagy and Aladár Czitrovszky for providing access to the optical profilometer.

## Data availability

The datasets generated during and/or analysed during the current study are available from the corresponding author on reasonable request.